\documentclass[printer]{aa}
\usepackage[english]{babel}
\usepackage{txfonts}
\usepackage{natbib}
\usepackage{aas_macros}
\usepackage{graphicx}
\def\xmm {\emph{XMM-Newton}}
\def\cxo {\emph{Chandra}}
\def\src {PSR\,J1357$-$6429}
\def\flux {\mbox{erg cm$^{-2}$ s$^{-1}$}}
\def\lum {\mbox{erg s$^{-1}$}}
\def\nh {$N_{\rm H}$}
\begin{document}
\title{Discovery of X-ray emission from the young radio pulsar \src}
\author{P.~Esposito\inst{1,2}
    \and A.~Tiengo\inst{2}
    \and A.~De~Luca\inst{2}
    \and F.~Mattana\inst{2,3}}
 \institute{Universit\`a degli Studi di Pavia, Dipartimento di Fisica Nucleare e Teorica and INFN-Pavia, via Bassi 6, I-27100 Pavia, Italy
     \email{paoloesp@iasf-milano.inaf.it}
\and INAF - Istituto di Astrofisica Spaziale e Fisica Cosmica Milano, via Bassini 15, I-20133 Milano, Italy
\and Universit\`a degli Studi di Milano - Bicocca, Dipartimento di Fisica G.~Occhialini, p.za della Scienza 3, I-20126 Milano, Italy}
\date{Received / Accepted}
\abstract{We present the first X-ray detection of the very young pulsar \src\ (characteristic age of 7.3 kyr) using data from the \xmm\ and \cxo\ satellites. We find that the spectrum is well described by a power-law plus blackbody model, with photon index $\Gamma=1.4$ and blackbody temperature $k_BT=160$ eV. For the estimated distance of \mbox{2.5 kpc}, this corresponds to a 2--10 keV luminosity of $\sim$$1.2\times10^{32}$ \lum, thus the fraction of the spin-down energy channeled by \src\ into X-ray emission is one of the lowest observed. The \cxo\ data confirm the positional coincidence with the radio pulsar and allow to set an upper limit of $3\times10^{31}$ \lum\ on the 2--10 keV luminosity of a compact pulsar wind nebula. We do not detect any pulsed emission from the source and determine an upper limit of 30\% for the modulation amplitude of the X-ray emission at the radio frequency of the pulsar. 
 \keywords{stars: individual (PSR\,J1357$-$6429) -- stars: neutron -- X-rays: stars}}
\titlerunning{X-ray observations of \src}
\authorrunning{P.~Esposito et al.}
\maketitle
\section{Introduction}
X-ray observations of radio pulsars provide a powerful diagnostic of the energetics and emission mechanisms of rotation-powered neutron stars. Due to the magnetic dipole braking, a pulsar loses rotational kinetic energy at a rate $\dot{E}=4\pi^2I\dot{P}P^{-3}$, where $I$ is the moment of inertia of the neutron star, assumed to be 10$^{45}$ g cm$^2$, and $P$ is the rotation period. Though pulsars have traditionally been mostly studied at radio wavelengths, only a small fraction (10$^{-7}$ to 10$^{-5}$, {e.g., \citealt{tml93}) of the ``spin-down luminosity'' $\dot{E}$ emerges as radio pulsations. Rotation power can manifest itself in the X\,/\,$\gamma$-ray energy range as pulsed emission, or as nebular radiation produced by a relativistic wind of particles emitted by the neutron star. Residual heat of formation is also observed as soft X-ray emission from young neutron stars. Such thermal radiation, however, can also be produced as a result of reheating from internal or external sources. The growing list of observable X-ray emitting rotation-powered pulsars allows the study of the properties of the population as a whole. Young pulsars constitute a particularly interesting subset to investigate owing to their large spin-down luminosities (\mbox{$\gtrsim$10$^{36}$ \lum}).\\
\indent The discovery of \src\ during the Parkes multibeam survey of the Galactic plane (see \citealt{lorimer06} and references therein) is reported in \citet{camilo04}. The pulsar is located near the supernova remnant candidate G309.8$-$2.6 \citep{duncan97} for which no distance or age information is available. With a spin period of 166 ms, a period derivative of $3.6\times10^{-13}$ s s$^{-1}$, and a characteristic age \mbox{$\tau_c=P/2\dot{P}\simeq7300$ yr}, this pulsar stands out as one of the ten youngest Galactic radio pulsars known (see the ATNF Pulsar Catalogue\footnote{See \texttt{http://www.atnf.csiro.au/research/pulsar/psrcat}\,.}, \citealt{manchester05}). The other main properties of this source derived from the radio observations are the spin-down luminosity of \mbox{$3.1\times10^{36}$ \lum} and the surface magnetic field strength of \mbox{$7.8\times10^{12}$ G}, inferred under the assumption of pure magnetic dipole braking. Based on a dispersion measure of \mbox{$\sim$127 cm$^{-3}$ pc} \citep{camilo04}, a distance of $\sim$2.4 kpc is estimated, according to the Cordes-Lazio NE2001 Galactic Free Electron Density Model\footnote{See \texttt{http://rsd-www.nrl.navy.mil/7213/lazio/ne\_model} and references therein.}.\\
\indent Here we report the first detection of \src\ in the X-ray range using the \xmm\ observatory and we present its spectral properties in the 0.5--10 keV energy band. We also made use of two short \cxo\ observations to confirm the identification and to probe possible spatial extended emissions, taking advantage of the superb angular resolution of the \cxo\ telescope.
\section{\xmm\ observation and data analysis\label{analysis}} 
In this section we present the results  obtained with the EPIC instrument on board the \xmm\ X-ray observatory. EPIC consists  of two MOS  \citep{turner01} and one pn CCD detectors \citep{struder01} sensitive to photons with energies between 0.1 and 15 keV. All the data reduction was performed using the XMM-Newton Science Analysis Software\footnote{See \texttt{http://xmm.vilspa.esa.es/}\,.} (SAS version 7.0). The raw observation data files were processed using standard pipeline tasks (\texttt{epproc} for pn, \texttt{emproc} for MOS data). Response matrices and effective area files were generated with the  SAS tasks \texttt{rmfgen} and \texttt{arfgen}.\\
\indent The observation was carried out on 2005 August 17 and had a duration of 15 ks, yielding net exposure times of \mbox{11.7 ks} in the pn camera and 14.5 ks in the two MOSs. The pn and the MOSs were operated in Full Frame mode (time resolution: \mbox{73.4 ms} and 2.6 s, respectively) and mounted the medium thickness filter. 
\src\  is clearly detected in the pn and MOS images (see Figure\,\ref{epic}) at the radio pulsar position (Right ascension = $13^{\rm{h}}\,57^{\rm{m}}\,02.4^{\rm{s}}$, Declination = $-64\degr\,29'\,30.2''$ (epoch J2000.0); \citealt{camilo04}). 
\begin{figure}[h!]
\centering
\resizebox{\hsize}{!}{\includegraphics[angle=0]{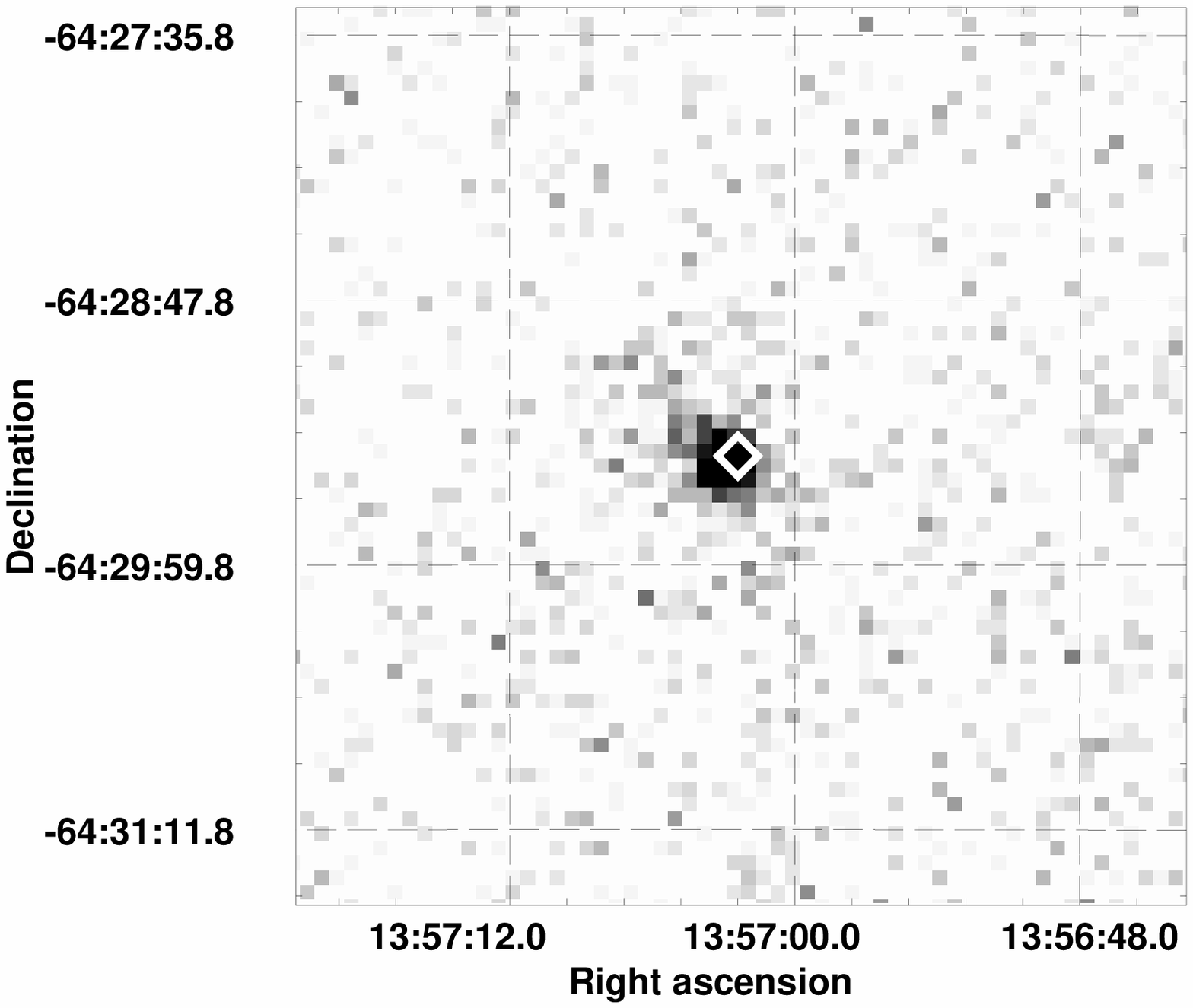}}
\caption{Field of \src\ as seen by the EPIC cameras in the 0.5--10 keV energy range. The radio pulsar position \citep{camilo04} is marked with the white diamond sign. The angular separation of the centroid of the X-ray source (computed using the SAS task \texttt{emldetect}) from the radio pulsar position is $(3.5\pm0.6)''$ (1\,$\sigma$ statistical error). Considering the \xmm\ absolute astrometric accuracy of $2''$ (r.m.s.), the X-ray and radio positions are consistent.}
\label{epic}
\end{figure}
The source spectra were extracted from circular regions centered at the position of \src. The whole observation was affected by a high particle background that led to the selection of a 20$''$ radius circle in order to increase the signal-to-noise ratio in the pn detector, particularly sensitive to particle background, and a 40$''$ radius for both the MOS cameras. The background spectra were extracted from annular regions with radii of 140$''$ and 220$''$ for the MOSs, and from two rectangular regions with total area of $\sim$$10^4$ arcsec$^2$ located on the sides of the source for the pn.
We carefully checked that the choice of different background extraction regions does not affect the spectral results. We selected events with pattern 0--4 and pattern 0--12 for the pn and the MOS, respectively. The resulting background subtracted count rates in the \mbox{0.5--10 keV} energy range were $(4.2\pm0.3)\times10^{-2}$ cts s$^{-1}$ in the pn and \mbox{$(1.9\pm0.2)\times10^{-2}$ cts s$^{-1}$} in the two MOS cameras, while the background rate expected in the source extraction regions is about 50\% of these values. The spectra were rebinned to have at least 20 counts in each energy bin. Spectral fits were performed using the \texttt{XSPEC} version 12.3 software\footnote{See \texttt{http://heasarc.gsfc.nasa.gov/docs/xanadu/xspec/}\,.}.\\
\indent The spectra from the three cameras were fitted together in the 0.5--10 keV energy range with a power law and with a power-law plus blackbody model (see Table\,\ref{fit}). The latter model provides a slightly better fit, with less structured residuals (see Figure\,\ref{spectra}). Furthermore, considering the distance of \mbox{2.5 kpc}, the interstellar absorption along the line of sight derived with the power-law fit is too low if compared to the typical column density of neutral absorbing gas  in that direction of
approximately \mbox{10$^{22}$ cm$^{-2}$} \citep{dickey90}. The resulting best-fit parameters for the power-law plus blackbody model are photon index $\Gamma=1.4$, blackbody temperature \mbox{$k_BT=0.16$ keV}, and absorption $N_{\rm H}=4\times10^{21}$ cm$^{-2}$ with a reduced $\chi^2$ of 0.85 for 70 degrees of freedom. The corresponding luminosity in the 0.5--10 keV band is $2.7\times10^{32}$ \lum.
\begin{table}
\begin{minipage}[h]{\columnwidth}
\caption{Summary of the \xmm\ spectral results. Errors are at the 90\% confidence level for a single interesting parameter.}
\label{fit}
\centering
\begin{tabular}{ccc}
\hline\hline
Parameter &  \multicolumn{2}{c}{Value}\\
 & PL & PL +BB \\
\hline
\nh\ (10$^{22}$ cm$^{-2}$) & $0.14^{+0.07}_{-0.06}$ & $0.4^{+0.3}_{-0.2}$\\
$\Gamma$ & $1.8^{+0.3}_{-0.2}$ & $1.4\pm0.5$\\
$k_BT$ (keV) & -- & $0.16^{+0.09}_{-0.04}$\\
R$_{BB}$\footnote{Radius at infinity assuming a distance of 2.5 kpc.} (km) & -- & $1.4^{+2.9}_{-0.2}$\\
Flux\footnote{Unabsorbed flux in the 0.5--10 keV energy range.} ($10^{-13}$\flux) & 2.3 & 3.6\\
Blackbody flux$^{b}$ ($10^{-13}$\flux) & -- & 1.3\\
$\chi^{2}_{r}$\,/\,d.o.f. & 1.00\,/\,72 & 0.85\,/\,70\\
\hline
\end{tabular}
\end{minipage}
\end{table} 
\begin{figure}[h!]
\resizebox{\hsize}{!}{\includegraphics[angle=-90]{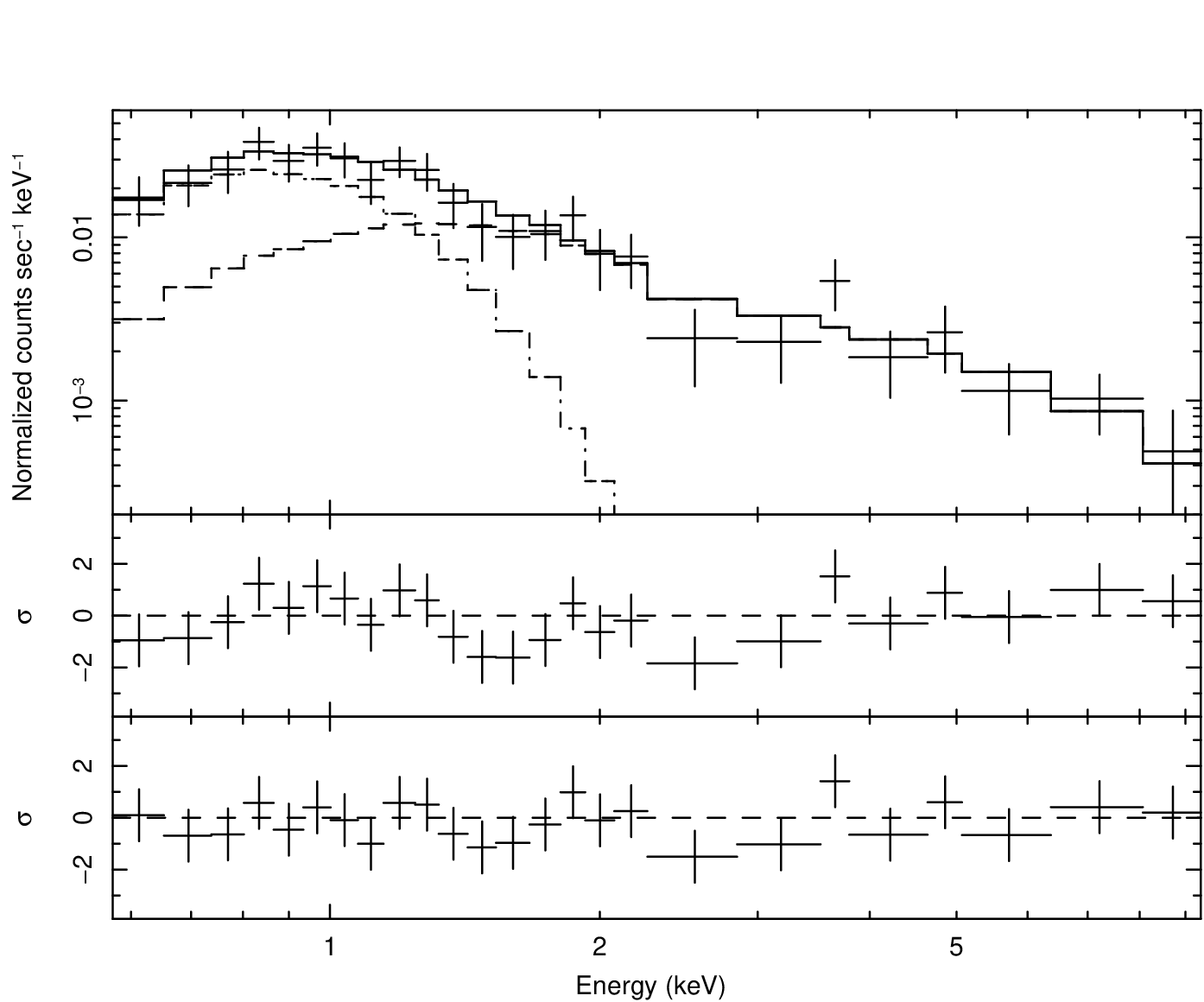}}
\caption{EPIC pn spectrum of \src. \emph{Top:} Data and best-fit power-law (dashed line) plus blackbody  (dot-dashed line) model.
 \emph{Middle:} Residuals from the power-law best-fit model in units of standard deviation.  \emph{Bottom:} Residuals from the power-law plus blackbody best-fit model in units of standard deviation.}
\label{spectra}
\end{figure}\\
\indent Young pulsars are often associated with pulsar wind nebulae: complex structures that arise from the interaction between the particle wind powered by the pulsar and the supernova ejecta or surrounding interstellar medium (see \citealt{gaensler06} for a review). 
Inspecting the EPIC images in various energy bands, we find only a marginal ($\approx$3\,$\sigma$) evidence of diffuse emission, in the 2--4 keV energy band consisting of a faint  elongated (\mbox{$\sim$20 arcsec} to the north-east, see Figure \ref{epic}) structure starting from \src. We took that excess as an upper limit for a diffuse emission: assuming the same spectrum as the point source, it corresponds to a 2--10 keV luminosity of \mbox{$\approx$$6\times10^{31}$ \lum}.\\
\indent For the timing analysis we applied the solar system barycenter correction to the photon arrival times with the SAS task \texttt{barycen}. We searched the data for pulsations around the spin frequency at the epoch of the \xmm\ observations, predicted assuming the pulse period  and the spin-down rate measured with the Parkes radio telescope \citep{camilo04}. As glitches and\,/\,or deviations from a linear spin-down may alter the period evolution, we searched over a wide period range centered at the value of $\sim$166 ms.
We searched for significant periodicities using two methods: a standard folding technique and the Rayleigh statistic. No pulsation were detected near to the predicted frequency with either method but, since the pn timing resolution (73 ms) allows to only poorly sample the \mbox{166 ms} pulsar period, a reliable upper limit on the pulsed fraction cannot be set.
\section{\cxo\ observations and data analysis\label{cxo-analysis}}
\src\ was observed by means of the \cxo\ \emph{\mbox{X-ray} Observatory} during two exposures of $\sim$17 ks duration each on 2005 November 18 and 19. The observations were carried out with the Spectroscopic array of the High Resolution Camera (HRC-S; \citealt{murray00}) used without transmission gratings. The HRC is a multichannel plate detector sensitive to X-ray over the \mbox{0.08--10 keV} energy range, although essentially no energy information on the detected photons is available. The HRC-S time resolution is 16 $\mu$s.\\
\indent We started from ``level 1'' event data calibrated and made available through the Chandra X-ray Center\footnote{See \texttt{http://cxc.harvard.edu/}.}. The level 1 event files contain all HRC triggers with the position information corrected for instrumental (degap) and aspect (dither) effects. After standard data processing with the Chandra Interactive Analysis of Observations (CIAO ver. 3.3), a point-like source has been clearly detected in both the observations at a position consistent with that of \src\ (see Figure\,\ref{psr}).
\begin{figure}[h!]
\centering
\resizebox{\hsize}{!}{\includegraphics[angle=0]{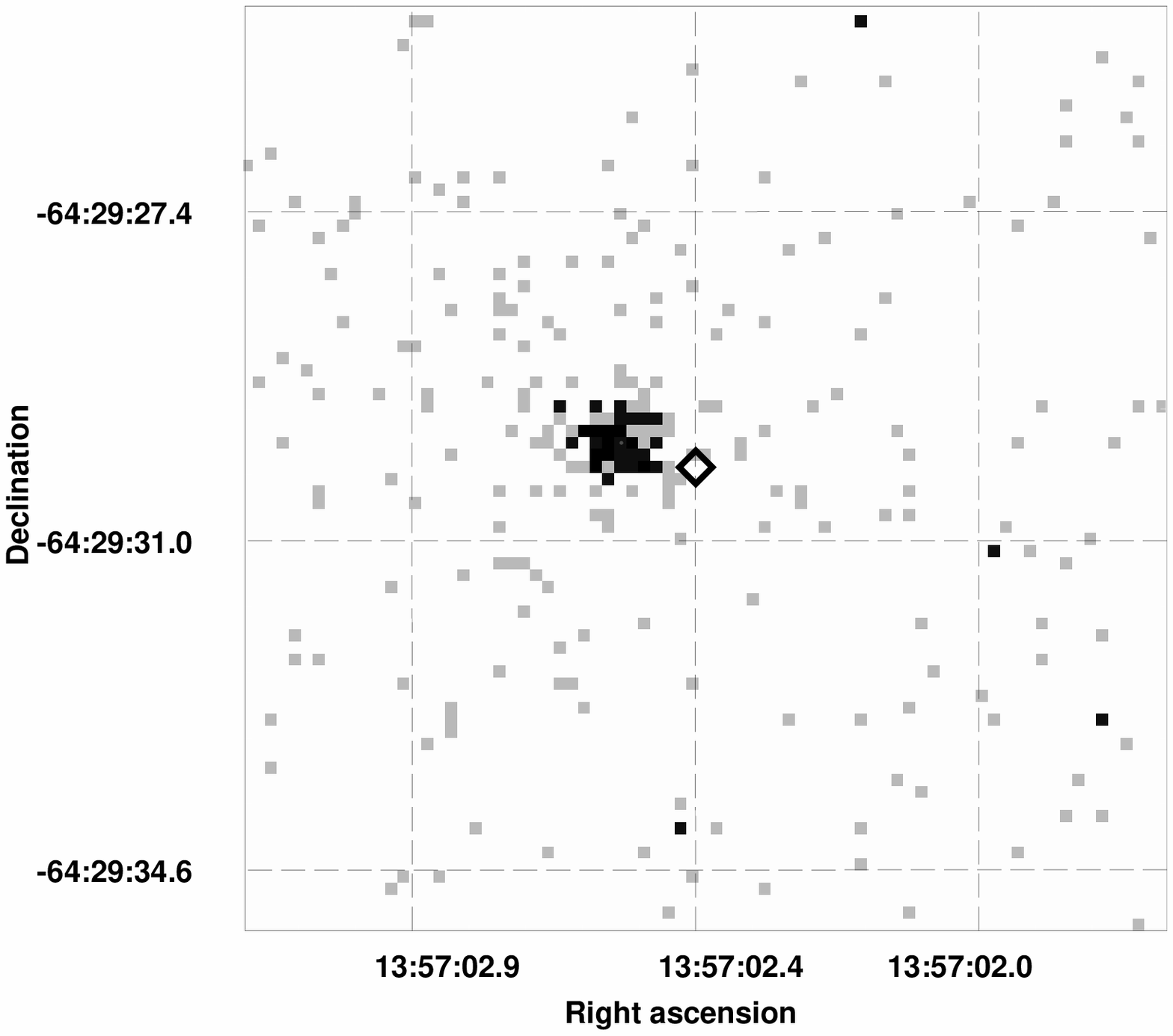}}
\caption{\cxo\ 0.08--10 keV HRC-S image centered on the radio pulsar position, marked with a diamond sign \citep{camilo04}. The CIAO \texttt{celldetect} routine yields a best-fit position for the X-ray source at an angular distance of $(0.9\pm0.2)''$ (1\,$\sigma$ statistical error) from the radio pulsar. This value consistent with the \cxo\ pointing accuracy of $0.8''$ (99\% confidence level).}
\label{psr}
\end{figure}\\
\indent For the timing analysis we corrected the data to  the solar system barycenter with the CIAO task \texttt{axbary} and then we followed the same procedure described in Section \ref{analysis}, but we again did not detect the source pulsation.
By folding the light curve of \src\ on the radio frequency and fitting it with a sinusoid, we determine a 90\% confidence level upper limit of $\sim$30\% on the amplitude of a sinusoidal modulation. We stress that this upper limit depends sensitively on data time binning and on the assumed pulse shape.\\
\indent We used the CIAO task \texttt{merge\_all} to generate a combined image of the source. 
Our main purpose was to search for diffuse structures on scales smaller than the \xmm\ angular resolution.
We compared the radial profile of the pulsar emission with the \cxo\ High-Resolution Mirror Assembly point-spread function at \mbox{1 keV} generated using Chandra Ray Tracer (ChaRT) and Model AXAF Response to X-rays (MARX). We found that the emission we detect from \src\ ($\sim$100 counts  concentrated within a $\sim$0.5$''$ radius circle) is consistent with that from a point source.\\
\indent We used the \cxo\ data and the PIMMS software\footnote{See \texttt{http://heasarc.gsfc.nasa.gov/docs/tools.html}\,.} to determine an upper limit on the luminosity of a possible spatial extended emission. 
The 3\,$\sigma$ upper limit on a pulsar wind nebula brightness (in counts s$^{-1}$) has been estimated as $3(bA)^{1/2}\tau^{-1}$, where $b$ is the background surface brightness in counts arcsec$^{-2}$, $A$ is the pulsar wind nebula area, and $\tau$ is the exposure duration. Assuming the interstellar absorption value from the \xmm\ best-fit model (\mbox{$N_{\rm H}=0.4\times10^{22}$ cm$^{-2}$}, see Section\,\ref{analysis}) and typical parameters for a pulsar wind nebula (radius of $\sim$$2\times10^{17}$ cm, that corresponds to $\sim$5$''$ for a distance of \mbox{2.5 kpc}, and power-law spectrum with photon index $\Gamma=1.6$, see, e.g., \citet{gotthelf03}), this upper limit corresponds to a 2--10 keV luminosity of \mbox{$\approx$$3\times10^{31}$ \lum} for a uniform diffuse nebula.
No significant diffuse excess was found even at larger angular scale, but the corresponding upper limit for diffuse emission is less constraining than that derived using the \xmm\ data.
\section{Discussion}
We have presented the results of the first X-ray observations of \src\ by means of the \xmm\ and \cxo\  observatories. The source has been positively detected in all the instruments although, probably due to the low statistics, we could not detect the source pulsation. The high angular resolution \cxo\ observations favor the picture in which most of the counts belong to a point source. We found that the spectrum is well represented by either a power-law with photon index $\Gamma=1.8^{+0.3}_{-0.2}$ or by a power-law plus blackbody model. In the latter case the best-fit parameters are for the power-law component a photon index $\Gamma=1.4\pm0.5$ and, for the blackbody component, radius\footnote{We indicate with $d_{\rm{N}}$ the distance in units of N kpc.} of $\sim$$1.4^{+2.9}_{-0.2}d_{2.5}$ km and temperature corresponding to $k_BT=0.16^{+0.09}_{-0.04}$ keV.\\
\indent It is generally believed that a combination of emission mechanisms are responsible for the detected X-ray flux from rotation-powered pulsars (see, e.g., \citealt{kaspi06} for a review). The acceleration of particles in the neutron star magnetosphere generates non thermal radiation by synchrotron and curvature radiation and\,/\,or inverse Compton processes, while soft thermal radiation could result by cooling of the surface of the neutron star. A harder thermal component can arise from polar-cap reheating, due to return currents from the outer gap or from close to the polar-cap. The dominant emission mechanism is likely related to the age of the pulsar. In pulsar younger than $\approx$10$^4$ yr the strong magnetospheric emission generally prevails over the thermal radiation, making difficult to detect it.\\
\indent As discussed in Section \ref{analysis}, we tend to prefer the power-law plus blackbody spectral model for \src.
 The resulting blackbody size of $\sim$1.5$d_{2.5}$ km may suggest that the soft emission ($\lesssim$2 keV) is coming from hot spots on the surface due to backflowing particles, rather than from the entire surface. However this hint should be considered with caution, as the surface temperature distribution of a neutron star is most likely non uniform (since the heath conductivity of the crust is higher along the magnetic field lines) and the small and hot blackbody could result from a more complicated distribution of temperature. Moreover, currently we lack of reliable models of cooling neutron star thermal emission and thus we cannot exclude that the soft component is emitted from surface layers of the whole neutron star.\\
\indent To date, thermal emission has been detected in only a few young radio pulsars. Among these, the properties of \src\ are similar to those {of the young pulsars Vela (PSR\,B0833$-$45; $\tau_c=11$ kyr, $P=89$ ms, \mbox{$\dot{E}=6.9\times10^{36}$ \lum}, and distance $d\simeq0.2$ kpc;  \citealt{pavlov01}) and PSR\,B1706--44 ( $\tau_c=17.5$ kyr, \mbox{$P=102$ ms}, \mbox{$\dot{E}=3.4\times10^{36}$ \lum}, and $d\simeq2.5$ kpc; \citealt{gotthelf02}). 
Notably, the efficiency in the conversion of the spin-down energy loss into X-ray luminosity for \src\ is $L_{\rm{0.5-10\,keV}}/\dot{E}\simeq8d_{2.5}^2\times10^{-5}$, significantly lower than the typical value of $\approx$$10^{-3}$ \citep{becker97}, and similar to that of PSR\,B1706--44 ($\sim$$10^{-4}$) and Vela ($\sim$$10^{-5}$).\\
\indent Although a pulsar wind nebula would not came as a surprise for this young and energetic source, we did not find clear evidence of diffuse X-ray emission associated with \src. However, some known examples of wind nebulae \citep[see][]{gaensler06}, rescaled to the distance of \src, would hide below the upper limits derived from the \xmm\ and \cxo\ data. \\
\indent New deeper exposures using \xmm\ or \cxo\ would help determine if a thermal component is present in the emission of \src\ as our spectral analysis suggests, and possibly detect a pulsed emission. High sensitivity observations would also serve to address the issue of the presence of a pulsar wind nebula. Although there is not any EGRET $\gamma$-ray source coincident with \src\ \citep{hartman99}, young neutron stars and their nebulae are often bright $\gamma$-ray sources and \src\ in particular, given its high ``spin-down flux'' $\dot{E}/d^2$ and similarity with Vela and PSR\,B1706--44, is likely to be a good target for the upcoming \emph{AGILE} and \emph{GLAST} satellites and the ground based Cherenkov air showers telescopes.

\begin{acknowledgements}
This work is based on data from observations with XMM-Newton, an ESA science mission with instruments and contributions directly funded by ESA member states and NASA. We also used data from the Chandra X-ray Observatory Center, which is operated by the Smithsonian Astrophysical Observatory Center on behalf of NASA. 
The authors thank the anonymous referee for helpful comments and acknowledge the support of the Italian Space Agency and the Italian Ministry for University and Research.
\end{acknowledgements}
\bibliographystyle{aa}
\bibliography{biblio}

\end{document}